\documentclass[prc,superscriptaddress,noshowpacs,unsortedaddress,twocolumn,showpacs,preprintnumbers,amsmath,amssymb]{revtex4-2}

\usepackage[dvipdfmx]{graphicx}
\usepackage{amsmath,amssymb,times}
\usepackage{color}
\usepackage{ulem}
\usepackage{bm}
\usepackage{here}

\def\lsim{~\,\makebox(1,1){$\stackrel{<}{\widetilde{}}$}\,~}

\newcommand{\beq}{\begin{equation}}
\newcommand{\eeq}{\end{equation}}
\newcommand{\bea}{\begin{eqnarray}}
\newcommand{\eea}{\end{eqnarray}}

\newcommand{\bfi}[1]{\mbox{\boldmath $#1$}}

\newcommand{\vK}{{\bfi K}}

\newcommand{\vs}{{\bfi s}}

\newcommand{\vrr}{{\bfi r}}
\newcommand{\vR}{{\bfi R}}

\def\a{\alpha}

\begin{document}
\title{ Neutron skin thickness of   $^{116,118,120,122,124}$Sn determined from reaction cross
    sections of proton scattering}

\author{Shingo~Tagami}
\affiliation{Department of Physics, Kyushu University, Fukuoka 819-0395, Japan}

  \author{Tomotsugu~Wakasa}
\affiliation{Department of Physics, Kyushu University, Fukuoka 819-0395, Japan}

\author{Masanobu Yahiro}
\email[]{orion093g@gmail.com}
\affiliation{Department of Physics, Kyushu University, Fukuoka 819-0395, Japan}             

\date{\today}

\begin{abstract}
\begin{description}
\item[Background]  
The cross sections of the isovector spin-dipole resonances (SDR) in the Sb isotopes
have been measured. Within the model used, the 
neutron-skin thicknesses $r_{\rm skin}({\rm exp})$ deduced  
$0.12 \pm 0.06$~fm for  $^{116}$Sn,
$0.13 \pm 0.06$~fm for $^{118}$Sn, 
$0.18 \pm 0.07$~fm for $^{120}$Sn, 
$0.22 \pm 0.07$~fm for  $^{122}$Sn, 
$0.19 \pm 0.07$~fm for  $^{124}$Sn. 
We tested the chiral (Kyushu) $g$-matrix folding model for 
$^{12}$C+$^{12}$C scattering, and found that 
the Kyushu $g$-matrix folding model is reliable for reaction cross sections 
$\sigma_{\rm R}$ 
in $30  \lsim E_{\rm in} \lsim 100 $~MeV and $250  \lsim E_{\rm in} \lsim 400$~MeV.  
We determine neutron skin thickness $r_{\rm skin}({\rm exp})$, using measured  $\sigma_{\rm R}$ of $^{4}$He+$^{116,120,224}$Sn scattering. The results are 
$r_{\rm skin}({\rm exp})=0.242 \pm 0.140$~fm for $^{116}$Sn, 
$r_{\rm skin}({\rm exp})=0.377 \pm 0.140$~fm for $^{120}$Sn,
$r_{\rm skin}({\rm exp})=0.180 \pm 0.142$~fm for $^{124}$Sn.
The $\sigma_{\rm R}$ are available for proton scattering on $^{116,118,120,122,124}$Sn 
with high accuracy of $2 \sim3\%$
as a function of incident energy $E_{\rm in}$. 
\item[Purpose]    
Our aim is to determine $r_{\rm skin}({\rm exp})$ for $^{116,118,120,122,124}$Sn with 
small errors by using the Kyushu $g$-matrix folding model. 
\item[Methods]    
Our model is the Kyushu $g$-matrix folding model  with the  densities scaled  from 
the D1S-GHFB+AMP neutron density, where D1S-GHFB+AMP stands for Gogny-D1S HFB (D1S-GHFB) with the angular momentum  projection (AMP). 
\item[Results]     
The proton radii of D1S-GHFB+AMP agree with those 
calculated with  the isotope shift method based on the electron scattering. 
We then scale the  D1S-GHFB+AMP neutron densities so as to reproduce the 
$\sigma_{\rm R}({\rm exp})$.  
In $30 \lsim E_{\rm in} \lsim 65$~MeV, 
we determine  $r_{\rm skin}({\rm exp})$ from measured $\sigma_{\rm R}$. 
The values are $r_{\rm skin}({\rm exp})=0.118 	\pm 0.021$~fm for $^{116}$Sn, 
$0.112 \pm 0.021$~fm for $^{118}$Sn, 
$0.124 \pm 0.021$~fm for $^{120}$Sn, 
$0.156 \pm 0.022$~fm for $^{124}$Sn. 
As for $^{122}$Sn, the skin value in $30 \lsim E_{\rm in} \lsim 50$~MeV is 
$0.122 \pm 0.024$~fm. 
\item[Conclusions] 
Our results are consistent with the previous values.
\end{description}
\end{abstract}

\maketitle


\section{Introduction and conclusion}
\label{Introduction}

{\it Background on experiments:} 
Horowitz, Pollock and Souder proposed a direct measurement 
for neutron-skin thickness $r_{\rm skin}=r_{\rm n}-r_{\rm p}$~\cite{PRC.63.025501}, 
where $r_{\rm p}$ and $r_{\rm n}$ are proton and neutron radii, respectively.  
This direct measurement  $r_{\rm skin}$ consists of parity-violating and elastic electron scattering. 
In fact, as for $^{208}$Pb, 
the PREX group has reported, 
\begin{equation}
r_{\rm skin}^{208}({\rm PREX2}) = 0.283\pm 0.071= 0.212 \sim 0.354
\,{\rm fm}, 
\label{Eq:Experimental constraint 208}
\end{equation}
combining the original Lead Radius EXperiment (PREX)  result \cite{PRL.108.112502,PRC.85.032501} 
with the updated PREX2 result \cite{Adhikari:2021phr}. This is the most reliable skin value 
for $^{208}$Pb.
Very lately, as for $^{48}$Ca, the CREX group has presented~\cite{CREX:2022kgg}. 
\bea
r_{\rm skin}^{48}({\rm CREX}) 
&=&0.121 \pm 0.026\ {\rm (exp)} \pm 0.024\ {\rm (model)}
\notag \\
&=&0.071\sim 0.171~{\rm fm} .  
\eea
The value is the most reliable skin value for $^{48}$Ca. 
These skin values and the $r_{\rm p}$ of Ref.~\cite{Angeli:2013epw,PRC.90.067304} 
allow us to deduce matter radii $r_{\rm m}$. These values are 
tabulated in Table \ref{reference values}.

As for the Sn isotopes, an indirect measurement on $r_{\rm skin}$ 
was made~\cite{Krasznahorkay:1999zz}. 
In 1998, the cross sections of the isovector spin-dipole resonances (SDR) in the Sb isotopes excited by the ($^{3}$He, t) charge-exchange reaction at 450 MeV for
$0^{\circ} \leq \theta_t \leq  1.15^{\circ}$ have been measured. 
 In order to deduce $r_{\rm n}$, they used the sum rule of Ref. \cite{Gaarde:1981vhv} valid 
for the spin-dipole operator involving the difference between the $\beta^{-}$ and $\beta^{-}$
strengths and the energy-weighted sum rule for the SDR calculated
in a model where the unperturbed particle-hole energies
are degenerate with an energy.
The skin values, $r_{\rm n}$, $r_{\rm m}$ and the $r_{\rm p}$ 
of Ref.~\cite{Angeli:2013epw} are also shown in Table \ref{reference values}.

As for  $^{120}$Sn, in 2018, the electric dipole strength distribution between 5 and 22 MeV was determined
at RCNP from polarization transfer observables measured in proton inelastic scattering at
$E_{\rm lab} = 295$~MeV and forward angles including $0^{\circ}$~\cite{Hashimoto:2015ema}. 
They extracted a highly precise electric dipole polarizability $\alpha_{\rm D} = 8.93(36)~{\rm fm}^3$ 
by combined it with photoabsorption data. 
Within the model used, they yield $r_{\rm skin}= 0.148(34)$~fm. Their results are also shown in 
 Table \ref{reference values}. The result has smaller error that that of Ref.~\cite{Krasznahorkay:1999zz}.

\begin{table}[htb]
\begin{center}
\caption
{Values of   $r_{\rm m}$,  $r_{\rm n}$, $r_{\rm skin}$, $r_{\rm p}$.  
The $r_{\rm p}$ are determined with the electron scattering, where the charge radii  are taken from Ref.~\cite{PRC.90.067304} for  $^{208}$Pb and Ref.~\cite{Angeli:2013epw} for  $^{48}$Ca and Sn isotopes. 
The radii are shown in units of fm.  
 }
\begin{tabular}{cccccc}
\hline\hline
 & Ref. & $r_{\rm p}$ & $r_{\rm m}$ &  $r_{\rm n}$ & $r_{\rm skin}$ \\
\hline
 $^{208}$Pb & PREX2 & $5.444$ & $5.617 \pm 0.044$ & $5.727 \pm 0.071$ & $0.283\pm 0.071$ \\
 $^{48}$Ca & CREX & $3.385$ & $3.456 \pm 0.050$ & $3.506 \pm 0.050$ & $0.121 \pm 0.050$ \\
  $^{116}$Sn & \cite{Krasznahorkay:1999zz} & $4.554$ & $4.67 \pm 0.06$ & $4.62 \pm 0.06$ & $0.12 \pm 0.06$ \\
  $^{118}$Sn & \cite{Krasznahorkay:1999zz} & $4.569$ & $4.70 \pm 0.06$ & $4.65 \pm 0.06$ & $0.13 \pm 0.06$ \\
  $^{120}$Sn & \cite{Krasznahorkay:1999zz} & $4.583$ & $4.76 \pm 0.07$ & $4.69 \pm 0.07$ & $0.18 \pm 0.07$ \\
  $^{120}$Sn & \cite{Hashimoto:2015ema} & $4.583$ & $4.731 \pm 0.034$ & $4.670 \pm 0.034$ & $0.148 \pm 0.034$ \\
  $^{122}$Sn & \cite{Krasznahorkay:1999zz} & $4.595$ & $4.82 \pm 0.07$ & $4.73 \pm 0.07$ & $0.22 \pm 0.07$ \\
  $^{124}$Sn & \cite{Krasznahorkay:1999zz} & $4.606$ & $4.80 \pm 0.07$ & $4.72 \pm 0.07$ & $0.19 \pm 0.07$ \\
\hline
\end{tabular}
 \label{reference values}
 \end{center} 
 \end{table}

{\it Background on model:} 
The reaction cross section $\sigma_{\rm R}$ is a standard way of determining matter radius  $r_{\rm m}$. 
One can  evaluate $r_{\rm skin}$ and $r_{\rm n}$ deduced from the $r_{\rm m}$ and the 
$r_{\rm p}$.~\cite{Angeli:2013epw} calculated with  the isotope shift method 
based on the electron scattering.

We tested the chiral (Kyushu) $g$-matrix folding model~\cite{Toyokawa:2017pdd} 
for $^{12}$C+$^{12}$C scattering and found that 
the Kyushu $g$-matrix folding model is reliable for reaction cross sections 
$\sigma_{\rm R}$ in 
$30  \lsim E_{\rm in} \lsim 100 $~MeV and $250  \lsim E_{\rm in} \lsim 400$~MeV~\cite{Tagami:2019svt}.  
The Kyushu $g$-matrix folding modeld were applied for measured  $\sigma_{\rm R}$ of $^{4}$He+$^{116,120,224}$Sn scattering~\cite{Matsuzaki:2021hdm}.; the results are 
$r_{\rm skin}({\rm exp})=0.242 \pm 0.140$~fm for $^{116}$Sn, 
$r_{\rm skin}({\rm exp})=0.377 \pm 0.140$~fm for $^{120}$Sn,
$r_{\rm skin}({\rm exp})=0.180 \pm 0.142$~fm for $^{124}$Sn.
These values have larger errors than those shown in Table \ref{reference values}.

As for $p$+$^{208}$Pb scattering, we determined a value of $r_{\rm skin}^{208}({\rm exp})$ 
from measured $\sigma_{\rm R}$ in a range of incident energies, 
$30 \lsim E_{\rm lab} \lsim 100$~MeV; 
the value is $r_{\rm skin}^{208}({\rm exp})=0.278 \pm 0.035$~fm~\cite{Tagami:2020bee}. Our result agrees with $r_{\rm skin}^{208}({\rm PREX2})$.
In this case, we used the D1S-GHFB+AMP proton and neutron densities, 
where D1S-GHFB+AMP stands for Gogny-D1S HFB (D1S-GHFB) 
with the angular momentum  projection (AMP). The $r_{\rm p}$ 
calculated with D1S-GHFB+AMP agirees with the experimental value of Ref.~\cite{PRC.90.067304}.

Also for  $^{116,118,120,122,124}$Sn, the $r_{\rm p}$ of D1S-GHFB+AMP agree with those~\cite{Angeli:2013epw} 
calculated with  the isotope shift method based on the electron scattering. 
For this reason, we use 
the D1S-GHFB+AMP proton and neutron densities in this paper.

The data~\cite{INGEMARSSON1999341,R.F.CARLSON:1995} on $\sigma_{\rm R}$ 
with high accuracy of $2 \sim3\%$ 
are available for p+$^{116,118,120,122,124}$Sn.  

{\it Aim:} 
Our aim is to determine $r_{\rm skin}({\rm exp})$ for $^{116,118,120,122,124}$Sn with 
small errors by 
using the Kyushu $g$-matrix folding model  with the D1S-GHFB+AMP proton and neutron   densities. 

{\it Results:} 
Our values are $r_{\rm skin}({\rm exp})=0.118 	\pm 0.021$~fm for $^{116}$Sn, 
$0.112 \pm 0.021$~fm for $^{118}$Sn, 
$0.124 \pm 0.021$~fm for $^{120}$Sn, 
$0.156 \pm 0.022$~fm for $^{124}$Sn, 
where the data are taken in $30 \lsim E_{\rm in} \lsim 65$~MeV. 
As for $^{122}$Sn, the skin value in $30 \lsim E_{\rm in} \lsim 50$~MeV is 
$0.122 \pm 0.024$~fm.

{\it Conclusion:} 
Our results of Table  \ref{result values} are consistent with those shown in Table \ref{reference values}.

\section{Model}
\label{Sec-Framework}

\begin{figure}[tbp]
\begin{center}
 \includegraphics[width=0.54\textwidth,clip]{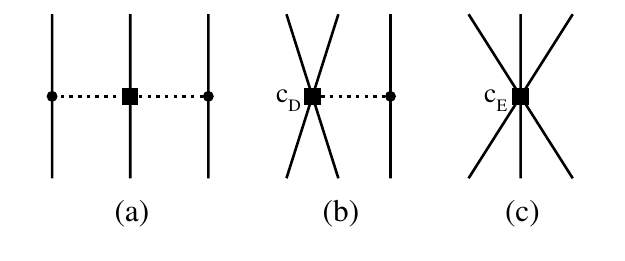}
 \caption{3NFs in NNLO. 
Diagram (a) corresponds 
to the Fujita-Miyazawa 2$\pi$-exchange 3NF \cite{PTP.17.360}, 
and diagrams (b) and (c) correspond to 1$\pi$-exchange and contact 3NFs.
The solid and dashed lines denote nucleon and pion propagations, 
respectively, and filled circles and squares stand for vertices. 
The strength of the filled-square vertex is often called $c_{D}$ 
in diagram (b) and $c_{E}$ in diagram (c). 
}
 \label{fig:diagram}
\end{center}
\end{figure}

Kohno calculated the $g$ matrix  for the symmetric nuclear matter, 
using the Brueckner-Hartree-Fock method with chiral N$^{3}$LO 2NFs and NNLO 3NFs~\cite{PRC.88.064005}. 
He set $c_D=-2.5$ and $c_E=0.25$ so that  the energy per nucleon can  become minimum 
at $\rho = \rho_{0}$; see Fig.~\ref{fig:diagram} for $c_{D}$ and $c_{E}$.
Toyokawa {\it et al.} localized the non-local chiral  $g$ matrix into three-range Gaussian forms~\cite{Toyokawa:2017pdd}, using the localization method proposed 
by the Melbourne group~\cite{von-Geramb-1991,Amos-1994}. 
The resulting local  $g$ matrix is called  ``Kyushu  $g$-matrix''.

Now, we show the folding model for nucleon-nucleus scattering. 
The potential $U(\vR)$ consists of the direct and exchange parts~\cite{PRC.89.064611},
$U^{\rm DR}(\vR)$ and $U^{\rm EX}(\vR)$, defined by 
\begin{subequations}
\begin{eqnarray}
U^{\rm DR}(\vR) & = & 
\sum_{\mu,\nu}\int             \rho^{\nu}_{\rm T}(\vrr_{\rm T})
            g^{\rm DR}_{\mu\nu}(s;\rho_{\mu\nu})  d
	    \vrr_{\rm T}\ ,\label{eq:UD} \\
U^{\rm EX}(\vR) & = & 
\sum_{\mu,\nu}
\int \rho^{\nu}_{\rm T}(\vrr_{\rm T},\vrr_{\rm T}+\vs) \nonumber \\
                &   &
\times g^{\rm EX}_{\mu\nu}(s;\rho_{\mu\nu}) \exp{[-i\vK(\vR) \cdot \vs/M]}
             d \vrr_{\rm T}\ ,\label{eq:UEX}
\end{eqnarray}
\end{subequations}
where $\vR$ is the coordinate between a projectile (P)  and 
a target (${\rm T}$),
$\vs=-\vrr_{\rm T}+\vR$, and $\vrr_{\rm T}$ is
the coordinate of the interacting nucleon from the center-of-mass of T.
Each of $\mu$ and $\nu$ denotes the $z$-component of isospin, i.e., 
$(1/2,-1/2)$ corresponds to (neutron, proton).
 The nonlocal $U^{\rm EX}$ has been localized in Eq.~\eqref{eq:UEX}
with the local semi-classical approximation
\cite{NPA.291.299,*NPA.291.317,*NPA.297.206},
where \vK(\vR) is the local momentum between P and T, 
and $M= A/(1 +A)$ for the target mass number $A$;
see Ref.~\cite{Minomo:2009ds} for the validity of the localization.
 The direct and exchange parts, $g^{\rm DR}_{\mu\nu}$ and
$g^{\rm EX}_{\mu\nu}$, of the $g$-matrix depend on the local density
\bea
 \rho_{\mu\nu}=\sigma^{\mu} \rho^{\nu}_{\rm T}(\vrr_{\rm T}+\vs/2)
\label{local-density approximation}
\eea
at the midpoint of the interacting nucleon pair, where $\sigma^{\mu}$ having ${\mu}=-1/2$
is the Pauli matrix of an 
incident proton. As a way of taking the center-of-mass correction to 
the D1S-GHFB+AMP densities,  we use the method of Ref.~\cite{PRC.85.064613}, 
since the procedure is quite simple. 

The direct and exchange parts, $g^{\rm DR}_{\mu\nu}$ and 
$g^{\rm EX}_{\mu\nu}$, of the $g$-matrix,  are described by~\cite{PRC.85.064613}
\begin{align}
&\hspace*{0.5cm} g_{\mu\nu}^{\rm DR}(s;\rho_{\mu\nu}) \nonumber \\ 
&=
\begin{cases}
\displaystyle{\frac{1}{4} \sum_S} \hat{S}^2 g_{\mu\nu}^{S1}
 (s;\rho_{\mu\nu}) \hspace*{0.42cm} ; \hspace*{0.2cm} 
 {\rm for} \hspace*{0.1cm} \mu+\nu = \pm 1 
 \vspace*{0.2cm}\\
\displaystyle{\frac{1}{8} \sum_{S,T}} 
\hat{S}^2 g_{\mu\nu}^{ST}(s;\rho_{\mu\nu}), 
\hspace*{0.2cm} ; \hspace*{0.2cm} 
{\rm for} \hspace*{0.1cm} \mu+\nu = 0 
\end{cases}
\\
&\hspace*{0.5cm}
g_{\mu\nu}^{\rm EX}(s;\rho_{\mu\nu}) \nonumber \\
&=
\begin{cases}
\displaystyle{\frac{1}{4} \sum_S} (-1)^{S+1} 
\hat{S}^2 g_{\mu\nu}^{S1} (s;\rho_{\mu\nu}) 
\hspace*{0.34cm} ; \hspace*{0.2cm} 
{\rm for} \hspace*{0.1cm} \mu+\nu = \pm 1 \vspace*{0.2cm}\\
\displaystyle{\frac{1}{8} \sum_{S,T}} (-1)^{S+T} 
\hat{S}^2 g_{\mu\nu}^{ST}(s;\rho_{\mu\nu}) 
\hspace*{0.2cm} ; \hspace*{0.2cm}
{\rm for} \hspace*{0.1cm} \mu+\nu = 0 ~~~~~
\end{cases}
\end{align}
where $\hat{S} = {\sqrt {2S+1}}$ and $g_{\mu\nu}^{ST}$ are 
the spin-isospin components of the $g$-matrix; see Ref.~\cite{PTEP.2018.023D03} for the explicit form of 
$g^{\rm DR}_{\mu\nu}$ and $g^{\rm EX}_{\mu\nu}$. 

As for Sn isotopes, the proton and neutron densities, $\rho_{\rm p}(r)$ and $\rho_{\rm n}(r)$, are calculated with D1S-GHFB+AMP~\cite{Tagami:2019svt}. 
As a way of taking the center-of-mass correction to the D1S-GHFB+AMP densities, 
we use the method of Ref.~\cite{Sumi:2012fr}, since the procedure is quite simple.

\subsection{Scaling procedure of neutron density}

The neutron density $\rho_p(r)$ is scaled from the D1S-GHFB+AMP one. 
We can obtain the scaled density $\rho_{\rm scaling}(\vrr)$ from the original density $\rho(\vrr)$ as
\bea
\rho_{\rm scaling}(\vrr)=\frac{1}{\a^3}\rho(\vrr/\a)
\eea
with a scaling factor
\bea
\a=\sqrt{ \frac{\langle \vrr^2 \rangle_{\rm scaling}}{\langle \vrr^2 \rangle}} .
\eea

We scale the neutron density so that  the 
$f \times \sigma_{\rm R}({\rm D1S})$ may reproduce the data ($\sigma_{\rm R}({\rm exp})$), 
where $\sigma_{\rm R}({\rm D1S})$ is the result of  D1S-GHFB+AMP and 
$f$ is the average of 
$\sigma_{\rm R}({\rm exp})/\sigma_{\rm R}({\rm D1S})$ over $E_{\rm lab}$.

\section{Results}
\label{Results}

\begin{figure}[H]
\begin{center}
 \includegraphics[width=0.5\textwidth,clip]{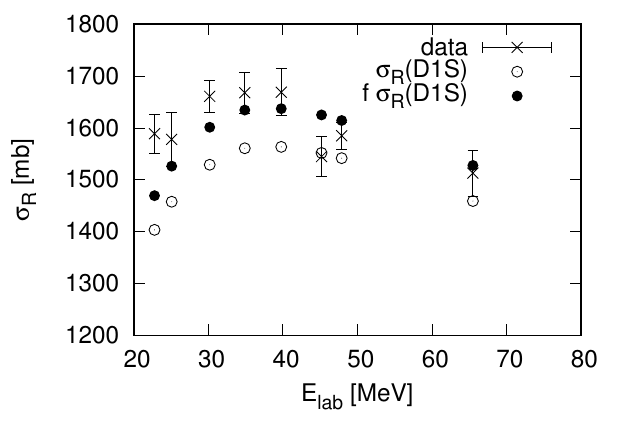}
 \caption{ 
 $E_{\rm lab}$ dependence of reaction cross sections $\sigma_{\rm R}$ 
 for $p$+$^{120}$Sn scattering. 
 Open circles stand for the results of  D1S-GHFB+AMP. 
 Closed circles correspond to $f \times \sigma_{\rm R}({\rm D1S}) $. 
 The data are taken from Refs.~\cite{INGEMARSSON1999341,R.F.CARLSON:1995}. 
   }
 \label{Fig-RXsec-p+120Sn}
\end{center}
\end{figure}

Figure~\ref{Fig-RXsec-p+120Sn} shows  reaction cross sections 
$\sigma_{\rm R}$ for p+$^{120}$Sn scattering as a function of $E_{\rm lab}$.
The $\sigma_{\rm R}({\rm D1S}) $ calculated wth  D1S-GHFB+AMP 
undershoots  the data~\cite{INGEMARSSON1999341,R.F.CARLSON:1995}
 ($\sigma_{\rm R}({\rm exp}) $) 
in $30.2 \leq E_{\rm lab} \leq 65.5$~MeV, but $f \times \sigma_{\rm R}({\rm D1S}) $ almost agrees 
with the data within error bars, where $f$ is the average of 
$f(E_{\rm lab}) \equiv \sigma_{\rm R}({\rm exp})/\sigma_{\rm R}({\rm D1S})$ over $E_{\rm lab}$. 
In this case, $f$ is 1.04711.
As a result of the scaling procedure mentioned above, 
we can obtain $r_{\rm m}=4.655 \pm 0.021$~fm, leading to $r_{\rm skin}({\rm exp})=0.124 \pm 0.021$~fm; see Table \ref{result values}.

\begin{figure}[H]
\begin{center}
 \includegraphics[width=0.5\textwidth,clip]{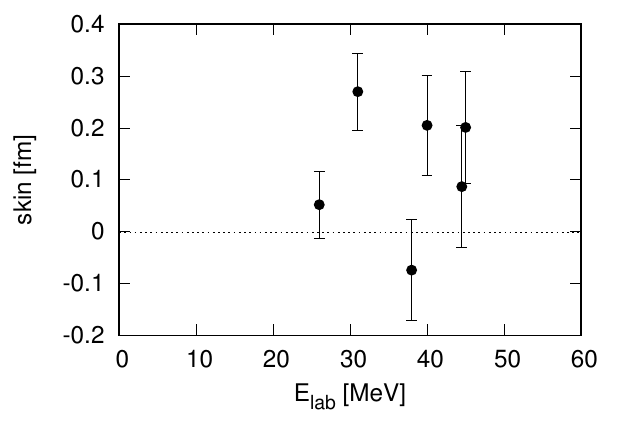}
 \caption{ 
$r_{\rm skin}({\rm exp})$ for each $E_{\rm lab}$ for  $p+^{120}$Sn scattering. 
 Closed circles with error-bar show  $r_{\rm skin}({\rm exp})$ for each $E_{\rm lab}$  
 }
 \label{Fig-RXsec-p+120Sn-e-skin}
\end{center}
\end{figure}

Figure~\ref{Fig-RXsec-p+120Sn-e-skin} shows a skin vale 
$r_{\rm skin}(E_{\rm lab}) $ for each $E_{\rm lab}$
 for $p$+$^{120}$Sn scattering in  
$30.2 \leq E_{\rm lab} \leq 65.5$~MeV 
The $r_{\rm skin}(E_{\rm lab}) $ fluctuate within 0.3~fm and -0.1~fm.
The indicates that taking the weighted mean is important.

\begin{figure}[H]
\begin{center}
 \includegraphics[width=0.5\textwidth,clip]{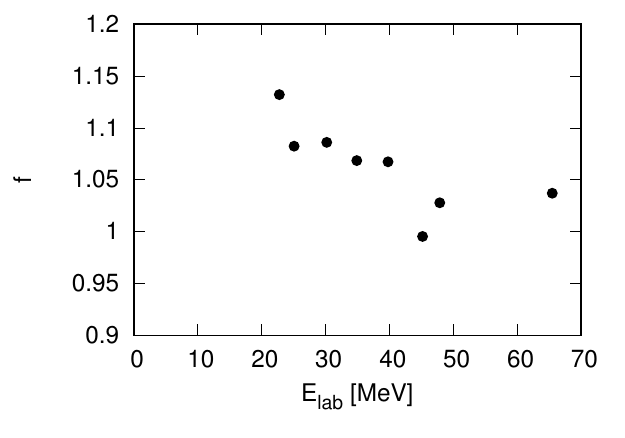}
 \caption{ 
 $E_{\rm lab}$ dependence of $f(E_{\rm lab})$ for  $p+^{120}$Sn scattering. 
 Closed circles show   $E_{\rm lab}$  of $f$.  }
 \label{Fig-RXsec-p+120Sn-F}
\end{center}
\end{figure}

Figure~\ref{Fig-RXsec-p+120Sn-F} shows  $E_{\rm lab}$ dependence of $f(E_{\rm lab})$ for $p$+$^{120}$Sn scattering. The $E_{\rm lab}$ dependence of $f(E_{\rm lab})$ is not smooth, 
because the $\sigma_{\rm R}$ calculated with D1S-GHFB+AMP are smooth 
for   $E_{\rm lab}$ dependence but the central values of  the data  are not. 
Note that the factor $f=1.04711$ is obtained by averaging $f(E_{\rm lab})$ over 
$30.2 \leq E_{\rm lab} \leq 65.5$~MeV

The same procedure is taken for p+$^{116,118,122,124}$Sn scattering. 
Our results and $f$ are shown in Table \ref{result values}. 
The $r_{\rm p}$ of  D1S-GHFB+AMP agree with those of the electron scattering, where the charge radii  are taken from Ref.~\cite{Angeli:2013epw}. 
The values of  $r_{\rm p}$ are shown in Table \ref{reference values}.

\squeezetable
\begin{table}[htb]
\begin{center}
\caption
{Values of  $f$, $r_{\rm m}$,  $r_{\rm n}$, $r_{\rm skin}$.  
The values of  $r_{\rm p}$ are shown in Table \ref{reference values}. 
The radii are shown in units of fm.  
 }
\begin{tabular}{cccccc}
\hline\hline
 & Ref. of data & $f$ & $r_{\rm m}$ &  $r_{\rm n}$ & $r_{\rm skin}$ \\
\hline
 $^{116}$Sn & \cite{INGEMARSSON1999341,R.F.CARLSON:1995} & $1.02447$ & $4.622 
 \pm 0.021$ & $4.672 \pm 0.021$ & $0.118 \pm 0.021$ \\
  $^{118}$Sn & \cite{INGEMARSSON1999341,R.F.CARLSON:1995} & $1.05118$ & $4.634 
 \pm 0.021$ & $4.681 \pm 0.021$ & $0.112 \pm 0.021$ \\
  $^{120}$Sn & \cite{INGEMARSSON1999341,R.F.CARLSON:1995}& $1.04711$ & $4.655 
 \pm 0.021$ & $4.706 \pm 0.021$ & $0.124 \pm 0.021$ \\
  $^{122}$Sn & \cite{R.F.CARLSON:1995} & $1.04881$ & $4.667 \pm 0.024$ & $4.717 
 \pm 0.024$ & $0.122 \pm 0.024$ \\
  $^{124}$Sn & \cite{INGEMARSSON1999341,R.F.CARLSON:1995} & $1.06002$ & $4.699 
 \pm 0.022$ & $4.761 \pm 0.022$ & $0.156 \pm 0.022$ \\
\hline
\end{tabular}
 \label{result values}
 \end{center} 
 \end{table}

\begin{figure}[H]
\begin{center}
 \includegraphics[width=0.5\textwidth,clip]{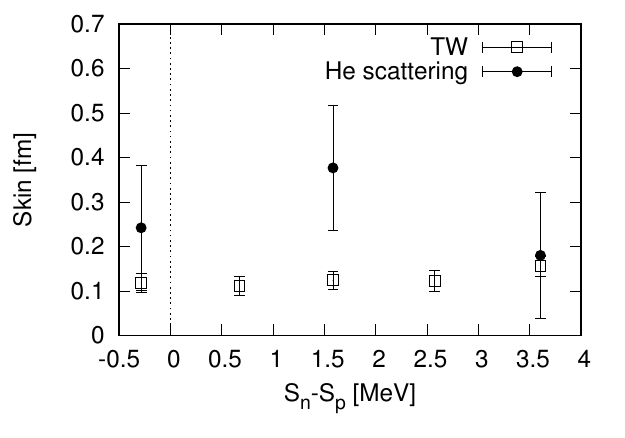}
 \caption{ 
 Skin values as a function of $S_{\rm p}-S_{\rm n}$. 
 Open squares stand for the results of this work (TW) for $^{116,118,120,122,124}$Sn. 
 The symbol ``$^{4}$He scattering'' stands for our previous work \cite{Matsuzaki:2021hdm} 
 for $^{4}$He scattering on $^{116,120,124}$Sn targets.
}
 \label{Fig-skins-compare}
\end{center}
\end{figure}

Figure \ref{Fig-skins-compare} shows skin values as a function of $S_{\rm p}-S_{\rm n}$. 
Our skin values calculated with D1S-GHFB+AMP are compared with 
our previous work \cite{Matsuzaki:2021hdm}  with Sly7, where the SLy7 parameter set is an improved
version of the widely used SLy4~\cite{Chabanat:1997un}. The data of 
measured $\sigma_{\rm R}$ for $^{4}$He scattering on $^{116,120,124}$Sn targets have larger errors than the data for $p$+$^{116,118,120,122,124}$Sn scattering. Eventually, our results have small errors compared with the previous results. This indicates that the present values are more reliable. 
As for  $^{120}$Sn, in addition, the present value 
$r_{\rm skin}=0.124 \pm 0.021$~fm 
 is consistent with $r_{\rm skin}=0.148 \pm 0.034$~fm~\cite{Hashimoto:2015ema} deduced from $\alpha_{\rm D}$. 
 Our values are near the lower bound of the previous result for $^{116}$Sn, and 
near the central value of the previous result for $^{124}$Sn.

Finally, we summarize skin values determined from measured $\sigma_{\rm R}$ and those by using  electroweak interaction.

\begin{figure}[H]
\begin{center}
 \includegraphics[width=0.5\textwidth,clip]{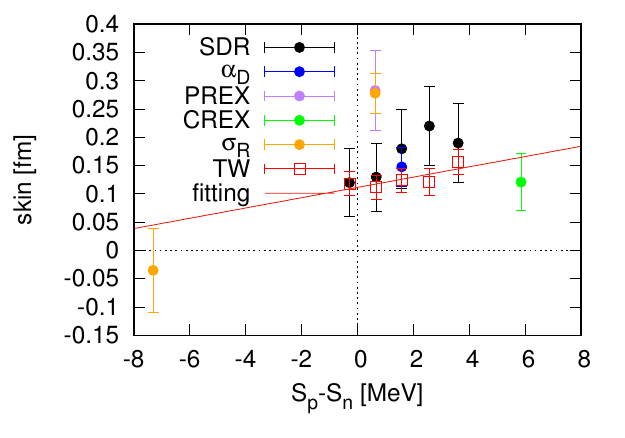}
 \caption{ 
 Skin values as a function of $S_{\rm p}-S_{\rm n}$. 
The symbol ``SDR" shows the results~\cite{Krasznahorkay:1999zz} of SDR in Sn isotopes. 
 The symbol `` $\a_D$'' means the results of the  $E1$ polarizability experiment ($E1$pE) 
for $^{120}$Sn~\cite{Hashimoto:2015ema}.  
 The symbol ``PREX'' stands for the result of $r_{\rm skin}^{208}({\rm PREX2})$, whereas the symbol ``CREX'' corresponds to  the result of $r_{\rm skin}^{48}({\rm CREX})$. 
 Open squares stand for the results of this work (TW) for $^{116,118,120,122,124}$Sn. 
 The symbol ``$\sigma_{\rm R}$'' stands for our previous works of Refs. \cite{Tagami:2020bee,Matsuzaki:2021hdm}. 
 The linear line shows $r_{\rm skin}=0.0091\delta+0.1116$. 
  The data ( closed circles with error bar) are taken from 
 Refs.~\cite{Krasznahorkay:1999zz,Hashimoto:2015ema,Adhikari:2021phr,CREX:2022kgg}.
}
 \label{Fig-skins}
\end{center}
\end{figure}

Figure \ref{Fig-skins} shows skin values as a function of $S_{\rm p}-S_{\rm n}$, 
where $S_{\rm p}$ ($S_{\rm n}$) is the proton (neutron) separation energy. 
The skin values  $r_{\rm skin}(\sigma_{\rm R})$ determined from measured $\sigma_{\rm R}$ 
for $^{116,118,120,122,124}$Sn are compared with the data of PREX2~\cite{Adhikari:2021phr}, 
$^{116,118,120,122,124}$Sn~\cite{Krasznahorkay:1999zz,Hashimoto:2015ema}, 
CREX~\cite{CREX:2022kgg}. As for Sn isotopes, 
our results of Table \ref{result values} are consistent with the previous experimental skin-values of Refs.~\cite{Krasznahorkay:1999zz,Hashimoto:2015ema}. 
Our value $r_{\rm skin}^{208}({\rm exp})=0.278 \pm 0.035$~fm of Ref.~\cite{Tagami:2020bee} agrees with $r_{\rm skin}^{208}({\rm PREX2})$.

Now we make qualitative discussion. 
We assume
a linear relation with $r_{\rm skin}$ and $\delta=S_{\rm p}-S_{\rm n}$ and take the $\chi^2$ fitting for 
our central skin-values for $^{116,118,120,122,124}$Sn, we 
can get $r_{\rm skin}=0.0091\delta+0.1116$. 
When we extrapolate our central skin-values for $^{116,118,120,122,124}$Sn by using 
the linear relation, we can obtain $r_{\rm skin}=0.165$~fm for  $^{48}$Ca.
In fact, we have already determined 
$r_{\rm skin}^{48}({\rm exp})=0.158 \pm	(0.023)_{\rm exp} \pm (0.012)_{\rm th}~{\rm fm}$
~\cite{TAGAMI2022105155} from $p$+$^{48}$Ca scattering 
and $^{48}$Ca+$^{12}$C scattering. 
These values are near the upper bound of CREX. 
As for  $^{40}$Ca, the linear relation yields $r_{\rm skin}^{48}({\rm exp})=0.045$~fm. The value  
is near the upper bound of  our previous value $r_{\rm skin}=-0.035 \pm 0.075$~fm~\cite{Matsuzaki:2021hdm} determined from  $^{4}$He+ $^{40}$Ca scattering. 
The skin values determined from  $\sigma_{\rm R}$ for $^{116,118,120,122,124}$, $^{40,48}$Ca 
are near the linear line; see the linear line of   Fog.~\ref{Fig-skins}.

\noindent
\begin{acknowledgments}
We would like to thank Toyokawa and Fukui for their contribution. 
\end{acknowledgments}




\bibliography{Folding-v17}

\end{document}